\documentclass[%
 reprint,
 nofootinbib,
 amsmath,amssymb,
 aps,
 prx,
]{revtex4-2}

\usepackage{lipsum}
\usepackage[capitalize]{cleveref}
\usepackage{upgreek}
\usepackage{graphicx}% Include figure files
\usepackage{dcolumn}% Align table columns on decimal point
\usepackage{bm}% bold math
\usepackage{xcolor}
\usepackage[normalem]{ulem}

\begin{document}

\preprint{APS/123-QED}

\title{Sedimentary rocks from Mediterranean drought in the Messinian age as a probe of the past cosmic ray flux}

\author{Lorenzo Caccianiga}
 \email{lorenzo.caccianiga@mi.infn.it}
\affiliation{%
 INFN - Sezione di Milano Via Celoria 16 20133 Milan, Italy
}%

\author{Lorenzo Apollonio}%
\email{lorenzo.apollonio@unimi.it}
  \affiliation{Dipartimento di fisica "A. Pontremoli" - Università degli Studi di Milano Via Celoria 16 20133 Milan, Italy}
  \altaffiliation[Also at ]{%
 INFN - Sezione di Milano
}%

\author{Claudio Galelli}%
  \affiliation{Laboratoire Univers et Théories, Observatoire de Paris, CNRS, 5 Place Jules Janssen 92190 Meudon}

\author{Federico Maria Mariani}%
\author{Paolo Magnani}%

  \affiliation{Dipartimento di fisica "A. Pontremoli" - Università degli Studi di Milano Via Celoria 16 20133 Milan, Italy}
  \altaffiliation[Also at ]{%
 INFN - Sezione di Milano
}%

\author{Alessandro Veutro}%
\affiliation{Università di Roma La Sapienza, I-00185 Roma, Italy\\}
  \altaffiliation[Also at ]{INFN - Sezione di Roma}

\date{\today}

\begin{abstract}
We propose the use of natural minerals as detectors to study the past flux of cosmic rays. This novel application of the \textit{paleo-detector} technique requires a specific approach as it needs samples that have been exposed to secondary cosmic rays for a well defined period of time. We suggest here the use of the evaporites formed during the desiccation of the Mediterranean sea ${\sim}6\,$Myr ago. These minerals have been created and exposed to the air or under a shallow water basin for ${\sim}500\,$kyr before being quickly submerged again by a km-scale overburden of water. We show that, by looking at the damages left in the minerals by muons in cosmic ray showers, we could detect differences in the primary cosmic ray flux during that period, as the ones expected from nearby supernova explosions, below the percent-level. We show also that little to no background from radioactive contamination and other astroparticles is expected for this kind of analysis.

\end{abstract}

\maketitle

\section{Introduction}
What happened in the past of our Galaxy and of our Solar System?  This question is still largely unanswered.
Electromagnetic waves are not helpful in this aspect, as they do not leave traces that can be recovered and measured millions or billions of years after they reach the Earth. Thus, we need a different \textit{messenger}. 

Cosmic rays are charged particles, mostly protons and nuclei, emitted in a number of non-thermal astrophysical processes. More than a century after their discovery, cosmic rays are now routinely detected. The spectrum of both the primary particles, the ones interacting with the upper atmosphere, and of their daughter secondary particles at sea level are well known. Cosmic ray emissions have been associated with extreme astrophysical events, such as supernovae \cite{FermiPionSN}. The search for specific isotopes produced in supernovae explosions, in particular ${}^{60}$Fe, in deep-sea deposits suggests the possibility of one or multiple near ($\sim 30-150\:$pc) and recent ($2-7\:$Myr) supernova explosions (see \cite{Ertel_2023} and references therein). These kinds of studies are however limited in time to a few Myr by the decay time of the isotopes and in space to a few hundreds of pc by the limited amount of materials ejected.

In this paper, we propose the investigation of the damages caused in natural minerals by secondary cosmic rays, i.e.\ the products of the interaction of primary cosmic rays in the Earth's atmosphere,  as a novel way to measure variations in the past flux of primary cosmic rays.
This could confirm, for example, the evidence for near-Earth supernovae mentioned before but also extend the search to a wider timescale and to different kinds of extreme galactic events. Aside from the astrophysical implications, there are several studies (see e.g. \cite{AtriCRLife,TsviGRBLife} and references therein) suggesting that such events could have an impact on the Earth's Biosphere, which could be investigated with this kind of analysis.

The search for linear defects left by particles in the crystalline structure of suitable minerals, called \textit{paleo-detectors}, has been proposed
in recent years, as a technique to detect dark matter~\cite{Baum:2018tfw,Drukier:2018pdy,Edwards:2018hcf} and neutrinos~\cite{Baum:2019fqm, tapia:2021}. These studies take advantage of the enormous exposure that can be acquired through age even with a small amount of material. 
Among neutrinos, atmospheric neutrino fluxes have been suggested as a proxy for the primary cosmic ray flux~\cite{Jordan:2020gxx}; their very small cross sections however make them unsuitable for detecting transients of duration  $\lesssim 100\,$Myr, as long integration periods are needed to collect enough statistics.

In a nutshell, the idea behind the paleo-detector technique is that particles interacting in a mineral cause nuclear recoils that damage permanently its structure. These linear defects, or \textit{tracks}, can range from a few nm to hundreds of $\upmu$m. The best technique to observe and count these tracks depends on the characteristics of the specific mineral and on the track lengths: for long tracks ($\upmu$m-scale) optical microscopes can be enough. Indeed, damages from spontaneous fission are routinely measured to date uranium-rich minerals. For shorter tracks, some different microscopy techniques have been suggested, such as Helium-Ion beam microscopy, small-angle X-ray scattering, and atomic force microscopy. All previous works on paleo-detectors focused on rare events, and as such considered secondary cosmic rays as the principal background to the signal. For this reason, candidate samples in these studies are planned to be dug up from km-scale depths underground. 

This work flips such an approach and suggests a strategy to employ paleo-detectors to investigate the past flux of cosmic rays and thus obtain information on the history of our Galaxy.

\section{paleo-detectors for cosmic ray flux measurements}
Using natural minerals as detectors has several advantages that have been previously outlined, but faces many challenges too. Firstly, the efficiency and speed of track detection depend on the size and nature of the damage left by the particles. These issues are shared with the other possible usages of paleo-detectors and are thus being tackled in a number of works that can be found referenced in~\cite{BAUM2023101245}. 
As an example, in a previous work concerning atmospheric neutrinos tracks~\cite{Jordan:2020gxx}, it is reported that imaging with small angle X-ray scattering at a synchrotron facility, a $\sim100\,\text{g}$ sample could be read out with a three-dimensional spatial resolution of $\sim15\,\text{nm}$.
Here we note that the tracks induced by cosmic rays can reach much larger sizes than the ones coming from dark matter or neutrino interaction, and are much more numerous. We then expect the challenge of detecting tracks induced by cosmic rays to be less complex than in the case of dark matter or neutrinos.

The other challenge in using natural minerals as particle detectors is the fact that they are naturally produced in uncontrolled environments and, if their structure is capable of recording damages caused by particles, they will do so as long as they are not destroyed or overheated, i.e.\ paleo-detectors cannot be turned off. This is hardly a problem if the purpose of using this kind of detector is extending the exposure, but it becomes one if we want to investigate the fluxes of astroparticles in the past. 

Cosmic rays, in this respect, are much more easily stopped by overburdens of material acting as a shield for the paleo-detector sample. If a sample has been exposed to the air, and thus to secondary cosmic rays, for a certain period of time, and then buried, the tracks in it preserve information on the flux during the exposure time. This principle is used also to evaluate the exposure time of rocks, in general measuring cosmogenic nuclides. 

\section{The Mediterranean desiccation}
Knowing the history of a mineral over the Myr scale is not trivial. We have identified a peculiar geological event that can act as an important case study in this respect: the Messinian salinity crisis. This event occurred ${\sim}6\,$Myr ago~\cite{MSCage}, when, under the most accepted scenario, the Mediterranean Sea desiccated following the closure of the precursor of the current Strait of Gibraltar. The evaporation of large parts of this huge water body led to the creation of many evaporites, mostly halite [NaCl] and gypsum [CaSO$_{4}$\,2(H$_{2}$O)], around ${\sim}5.6\,$Myr ago~\cite{MSCage, MSCev}. These minerals were exposed to the air, and thus to secondary cosmic rays, directly or under shallow water. After ${\sim}300\,$kyr, the connection to the Atlantic Ocean opened again and the Mediterranean basin was quickly ($O(\textrm{yr})$) filled. The huge inundation called the Zanclean flood, dragged many sediments to the deepest points of the basin, where they still are, covered by a $\sim\,$km-scale overburden of water. 

Halite has been suggested already as a potential paleo-detector~\cite{Baum:2018tfw,Jordan:2020gxx} and experimental tests are being performed to identify the best technique to measure tracks in this mineral~\cite{HEiKA}. The first observation of nuclear recoil tracks in halite was very recently reported \cite{mastersthesis-full}. The tracks were observed in a length range between ${\sim}1-1000\,$nm and the source of the nuclear recoils is still under study.

\par Halite was also the target of promising pioneering studies regarding the identification of tracks using the optical technique known as \textit{color centers}~\cite{Fowler:1975}. There is scarce information in the literature about track annealing in Halite, i.e. we have no information on the temperature at which tracks tend to fade away if the crystal is heated. However, typical anneal temperatures range from $\sim 70^\circ$C in plastic detectors to hundreds of degrees in minerals such as mica \cite{GUO2012233}. When compared to other applications of paleo-detectors previously considered, the halite samples we plan to retrieve are conserved in an environment that is expected to be more stable thermally and not subject to high-temperature peaks, if we discard sampling locations close to volcanic activities. In any case, the annealing process would be the same on background tracks, most notably on the ones coming from spontaneous fission. This would make it possible, in principle, to derive a track fading calibration based on the measured fission tracks and apply it to the signal. 

We note here that, while the desiccation of the Mediterranean Sea is globally accepted by the geological community, whether the deposition of evaporites occurred mostly in shallow waters or deeper ones (some hundreds of meters) is still debated and most likely dependent on the different locations. Similarly, the deposition rate can vary based on the assumed conditions (see e.g. \cite{MEILIJSON2019374} and references therein). It has even been suggested that the creation of this large amount of evaporites might have a different explanation, limiting the entity of the desiccation of the sea~\cite{Ryan2023}. A study of the cosmic ray-induced tracks in the evaporites from the depths of the Mediterranean could also help prove or disprove the different theories. 

\begin{figure*}[ht]
  \includegraphics[width=0.42\textwidth, height=5cm]{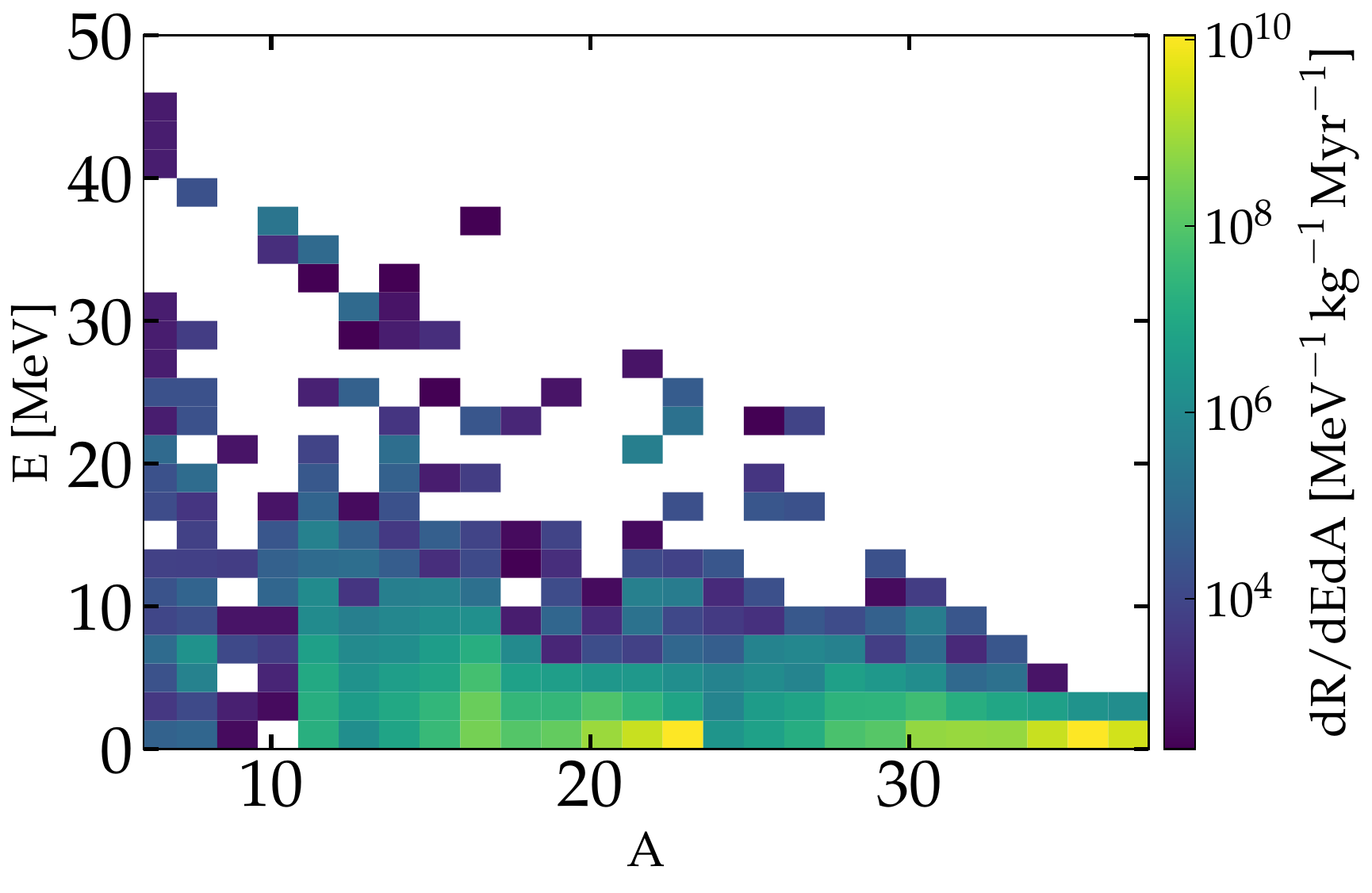}
  \includegraphics[width=0.42\textwidth, height=4.85cm]{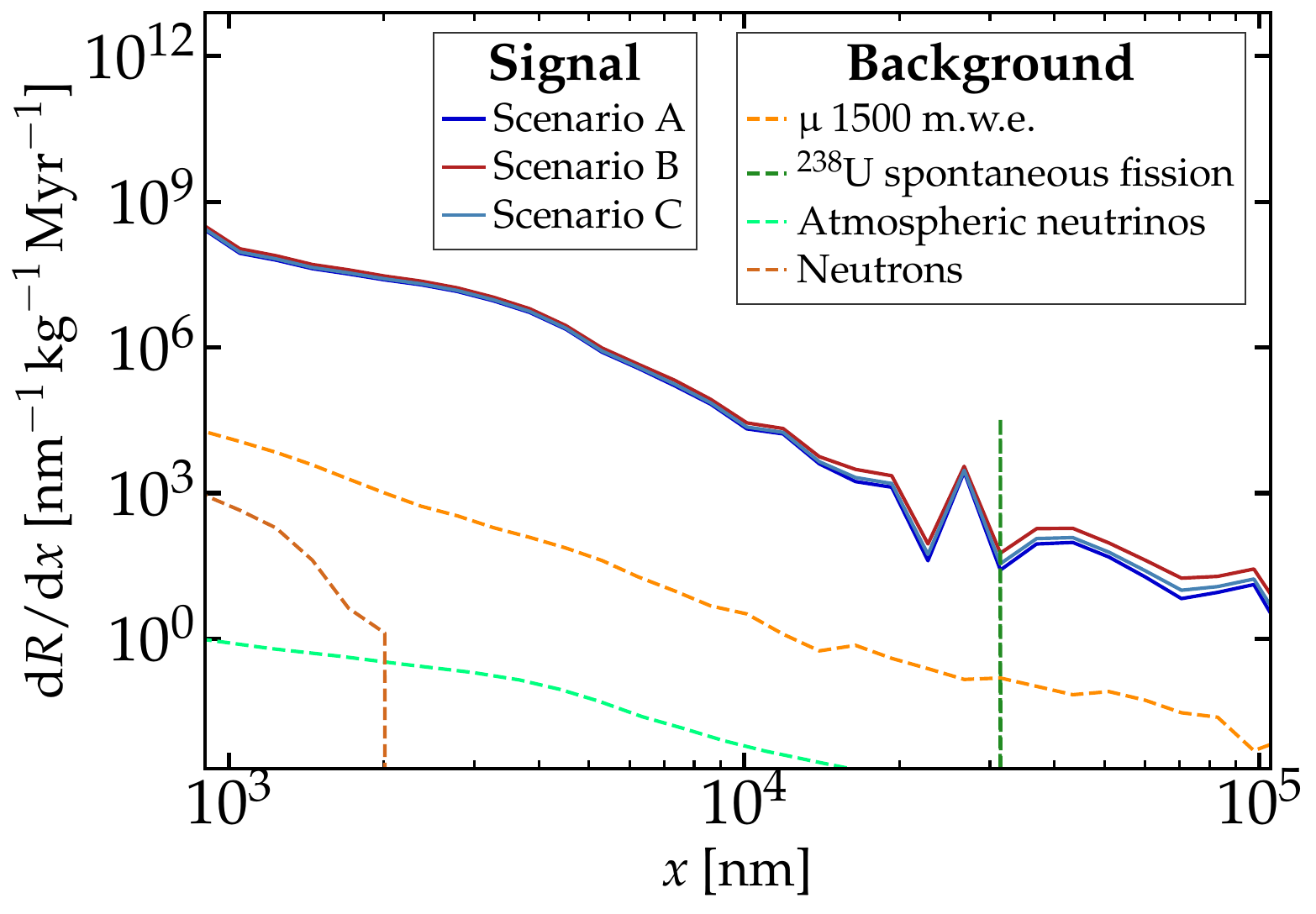}
  \caption{\label{fig:dRdx} Left: Double differential rate with respect to the nuclear recoil energy ($E$) and the atomic mass number ($A$) when a muon flux equal to the current one pass by the sample of halite. Right: Differential rate of track lengths. We show three cases as signals. Scenario A considers the current measured $\upmu$ flux at sea level. Scenario B considers the $\upmu$ flux coming from an SN distant 20\,pc, in addition to the current flux. Scenario C considers the explosion of an SN at 100\,pc. The backgrounds are represented by internal radioactivity (spontaneous fission and neutrons) and external astroparticles (atmospheric neutrinos). We included amongst the backgrounds the muon flux under 1.5 km of water which is used for computing the final number of tracks after a precise exposure history in \cref{fig:integral}}
\end{figure*}

\section{Simulation}
To simulate the number of tracks that can be expected in a halite sample produced during the Mediterranean desiccation, we used the python package \verb+Paleopy+ \footnote{https://github.com/tedwards2412/paleopy}, developed for the pioneering work describing the paleo-detectors technique~\cite{Edwards:2018hcf} and subsequently used in the following papers. The software considers the differential nuclear recoil rates, due to the interaction of the recoiling particle with the nuclei inside of the mineral, as a function of the recoil energy. The recoil rates are then converted into track rates ($\mathrm{d}R/\mathrm{d}x$) as a function of the track length ($x$) using the stopping powers of each nucleus, evaluated through the software \verb+SRIM+. 

The backgrounds are represented by other astroparticles and radioactive decay within the mineral or its immediate surroundings. From \verb+Paleopy+ we obtained the expected track length distribution for all the backgrounds with the exception of atmospheric neutrinos which are obtained directly from the dedicated studies in halite available in \cite{Jordan:2020gxx}.

$\alpha$ particles from radioactive decays do not deposit enough energy in the material to form tracks in most minerals, including halite~\cite{GUO2012233}. However, they can induce ($\alpha$,n) reactions, and the emitted neutrons are the most significant background for track lengths lower than few $\upmu$m. This background, which is the most important for DM and neutrino searches, is in any case at least 1 order of magnitude lower than the expected cosmic ray signal in the range of track lengths above a few hundred nanometers. The only background that is expected to be of the same order of magnitude as the signal is the one coming from the spontaneous fission of ${}^{238}$U. For this simulation, we consider a ${}^{238}$U concentration of $5\times10^{-6}\,\mathrm{g}/\mathrm{g}$.  The fission fragments, however, have a limited range of Z and energy and thus produce tracks of very specific lengths, for halite between ${\sim}20\,\upmu$m and ${\sim}40\,\upmu$m. The fission tracks could potentially be used as calibration and reference if the age and uranium content of the sample can be determined accurately enough. 

Regarding the signal, we considered three cases of study. In the first one (Scenario A), no supernova exploded near the Earth while the mineral was directly exposed to muons, and thus we considered the current measured muon flux at sea level (from~\cite{GaisserEngelResconi}). The second one (Scenario B) considers, in addition to the current flux, the one coming from a supernova distant 20$\,\mathrm{pc}$. In the third one (Scenario C), the supernova exploded at a distance of 100$\,\mathrm{pc}$. We obtained the expected flux at the ground from the nearby supernovas from the work published in \cite{Thomas_2023} and its supplementary material \cite{Thomas_Zenodo}. The flux was computed considering a supernova with a total energy of $2.5 \times 10^{51}\,\mathrm{erg}$ and a conversion efficiency of 10\%. We note here that in \cite{Thomas_2023} a reference flux in case of no supernova event is also computed. It matches the one reported in \cite{GaisserEngelResconi} at high energies but is slightly lower in the $\sim$GeV region. We decided to use the second one to be conservative.

We have simulated the signal by inputting the muon spectrum in each scenario in a \verb+Geant4+~\cite{GEANT4:2002zbu} simulation. We simulated the muon interactions occurring inside of a cylindrical volume of halite with radius $10\,$m and height $10\,$m. The halite sample is defined as a pure composition of Na and Cl, with isotope fractions selected according to the NIST database of Atomic Weights and Isotopic Compositions~\cite{NIST2015}.
We have then extracted the expected differential nuclear recoil rate, both for the constituent nuclei and the lighter nuclei produced by inelastic scatterings, and inputted the results into \verb+Paleopy+.

\section{Results and discussion}
The expected differential rate of tracks per unit target mass and unit time is shown in \cref{fig:dRdx}. The left plot shows the double differential distribution with respect to the recoil energy ($E$) and the atomic mass number of the nucleus ($A$), computed with the current muon flux at Earth. The right plot shows the differential distribution with respect to the recoil track length ($x$). Since the total number of muons produced by a supernova reaching the Earth is not steady in time, here we show the mean value over the exposure time, added to the current flux of muons at the ground. 

It is clear that, in this simulation, the signal is above the expected backgrounds at all track lengths with the exception of fission tracks, as discussed before. These results are in line with the request from other paleo-detector applications to retrieve samples from large depths to avoid any cosmic ray exposure that would overwhelm the signal from other astroparticles. 

It should be noted that \cref{fig:dRdx} shows an "instantaneous" spectrum of tracks when both signal and backgrounds are actively targeting the mineral. However, our scenario involves a limited period of exposure to the signal while the backgrounds continue leaving tracks in the mineral throughout its life. As a first, simplified simulation, we consider a case in which the sample is created at the beginning of the Messinian age, $5.6\,$Myr ago ~\cite{MSCage, MSCev} exposed for $270\;$kyr and then submerged. This does not take into account the continuous deposit of evaporites before the complete desiccation. In the search for transient events, this represents a worst-case scenario, as the long exposure reduces the effect of the short burst of cosmic rays on the total track count. We computed also the background due to muons reaching the sample while submerged by water (\textit{off} period), using $1.5\,$km as reference depth, which is the average depth of the Mediterranean Sea, showing also a dispersion band of $\pm 500\,$m. 
The analytical form of the suppression factor is taken from \cite{PDG} \footnote{see Eq.~(30.6) and (30.7) of \cite{PDG}}.
We note here that, given the brevity of the SN signal with respect to the length of the \textit{off} period and the exponential suppression factor due to the water overburden, a SN exploding even at a few tens of pc during the submerged phase would not increase significantly the background, thus not impacting the final results.

\begin{figure}[ht]
\includegraphics[width=\columnwidth]{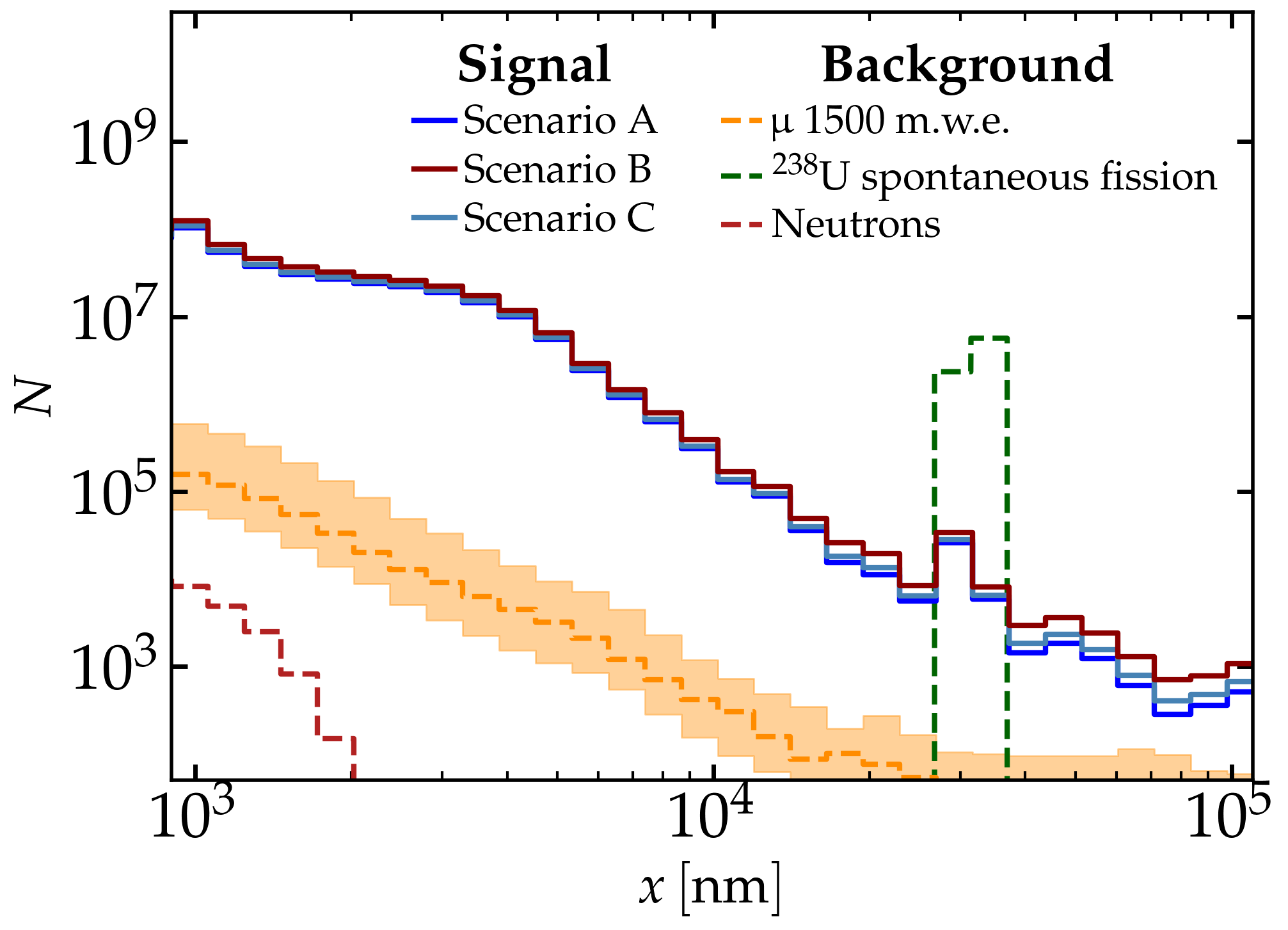} 
\includegraphics[width=\columnwidth]{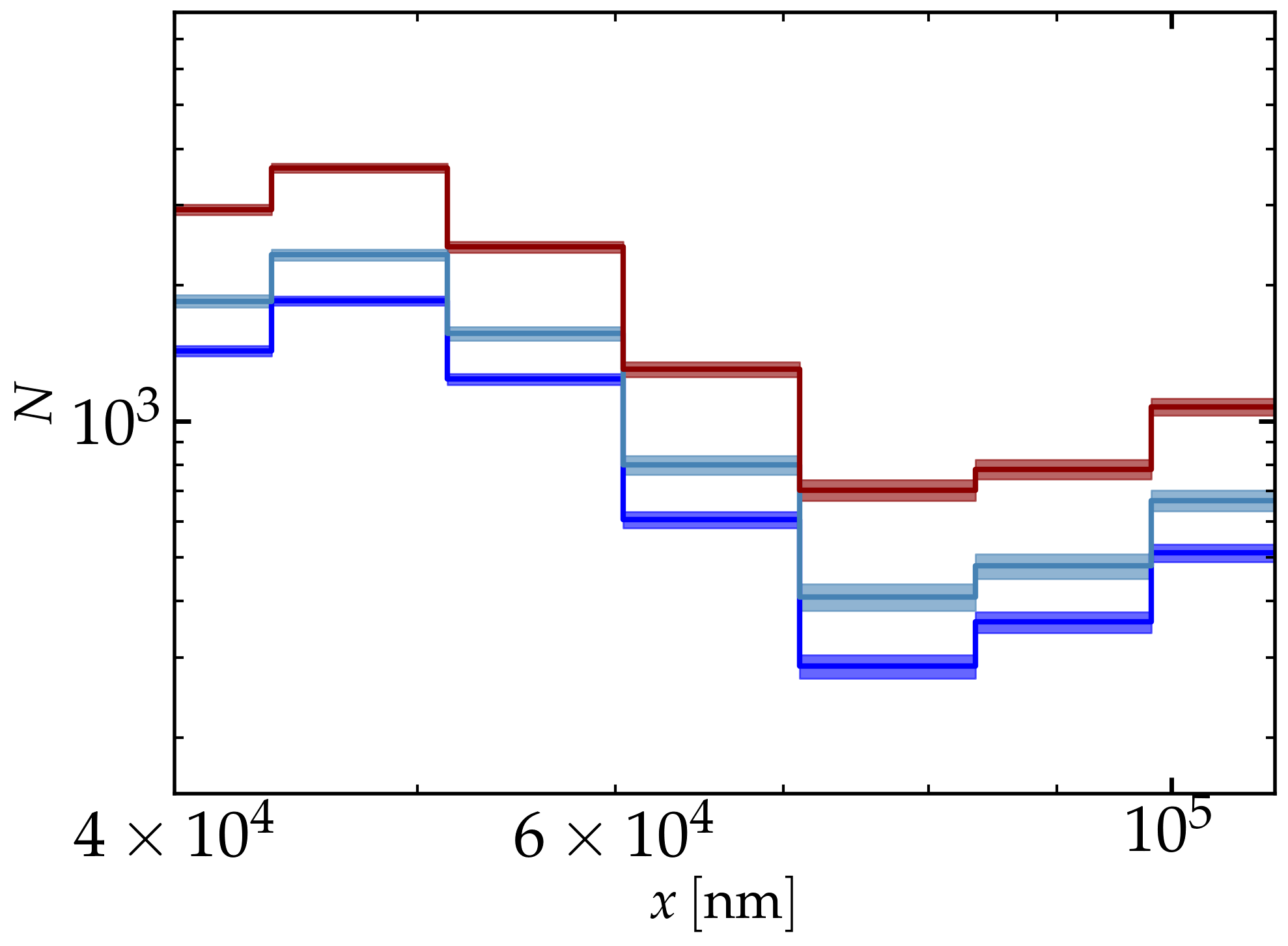} 
\caption{\label{fig:integral} Top: expected total number of tracks in a $10\,$g sample of halite which was created $5.6\,$Myr ago, exposed for $270\,$kyr to muons and then covered by a $1.5\,$km overburden of water. The backgrounds are integrated for the whole age of the sample with the exception of the underwater $\mu$, which is integrated for the period when the sample is sea-covered, i.e. $5330\,$kyr. Bottom: zoom of top plot in the track length region $40\,\upmu\mathrm{m}$-$100\,\upmu\mathrm{m}$.}
\end{figure}

Integrating the contributions on these different intervals we obtain the absolute number of tracks ($N$) shown in the top plot of \cref{fig:integral}.

We show the results for a $10\,$g sample. This mass was chosen as a very conservative limit, suggested also in many previous paleo-detectors feasibility studies~\cite{Edwards:2018hcf} where the signal tracks were much shorter (sub-$\upmu$m). As previously mentioned, in \cite{Jordan:2020gxx} the analysis of a $100\,$g Halite sample was envisioned to look for atmospheric neutrinos. No matter the chosen track detection method, the search for shorter tracks will always require more time and for this reason, we expect that higher volumes should be also within reach for the search for cosmic ray tracks.
We expect to be able to gather much larger masses of sample material (at least $\sim$kg), from which we could extract a number of $\sim 10\,$g pure enough crystals.

We can see that even with different exposures the signal in both scenarios is still expected to be well above all the backgrounds apart from the spontaneous fission peak. In general, all the internal backgrounds are much lower than signal. This implies that even large errors on the estimation of the age of the sample will not impact the final results. The tracks due to muons hitting the mineral while submerged, despite being the main background throughout the track length range, are always a few orders of magnitude below the signal tracks. This means that the final measurement is very poorly dependent on the muon flux variations during the \textit{off} period. It also opens the possibility to perform a similar search on older minerals, if similar suitable geological events are discovered.

The bottom plot of \cref{fig:integral} shows a zoom in the track length region $5\,\upmu\mathrm{m}$-$15\,\upmu\mathrm{m}$. This zoom highlights the region in which the two different signal scenarios are the most separated. Even with an additional, hypothetical counting error to $N$ up to $\sim 15\% (7\%)$ of the total number of tracks, we could still distinguish scenario B(C) from scenario A.

We also computed the minimal difference of cosmic ray flux between the Messinian age and now we would be able to distinguish. Under a simple approach of having the cosmic ray spectrum shape identical to the current one, with $10\,$g of target material, and just considering the uncertainties related to the Poissonian fluctuations on track formation, we could distinguish a normalization difference of $O(1\%)$.

\begin{figure}[t]
\includegraphics[width=\columnwidth]{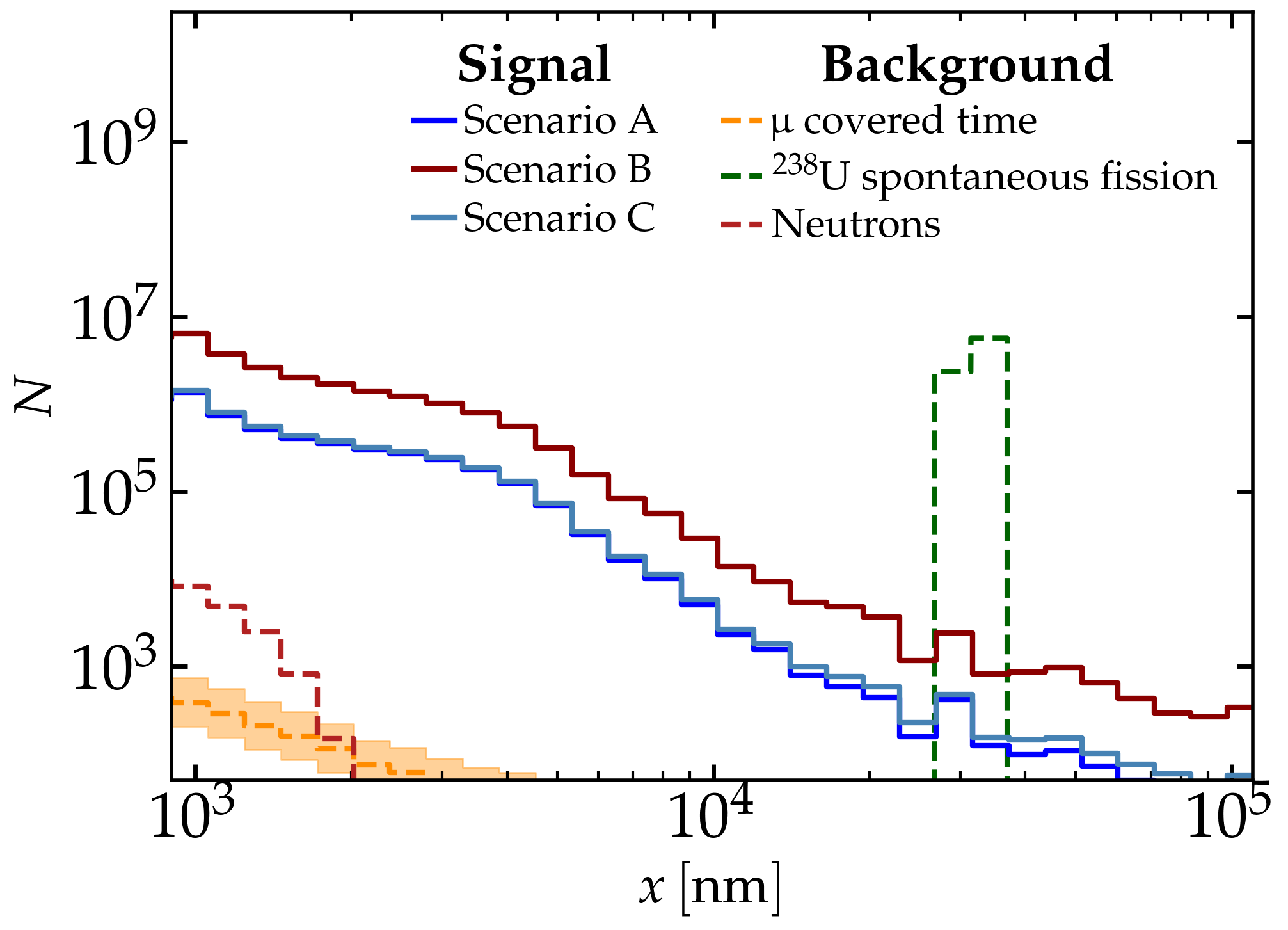} 
\includegraphics[width=\columnwidth]{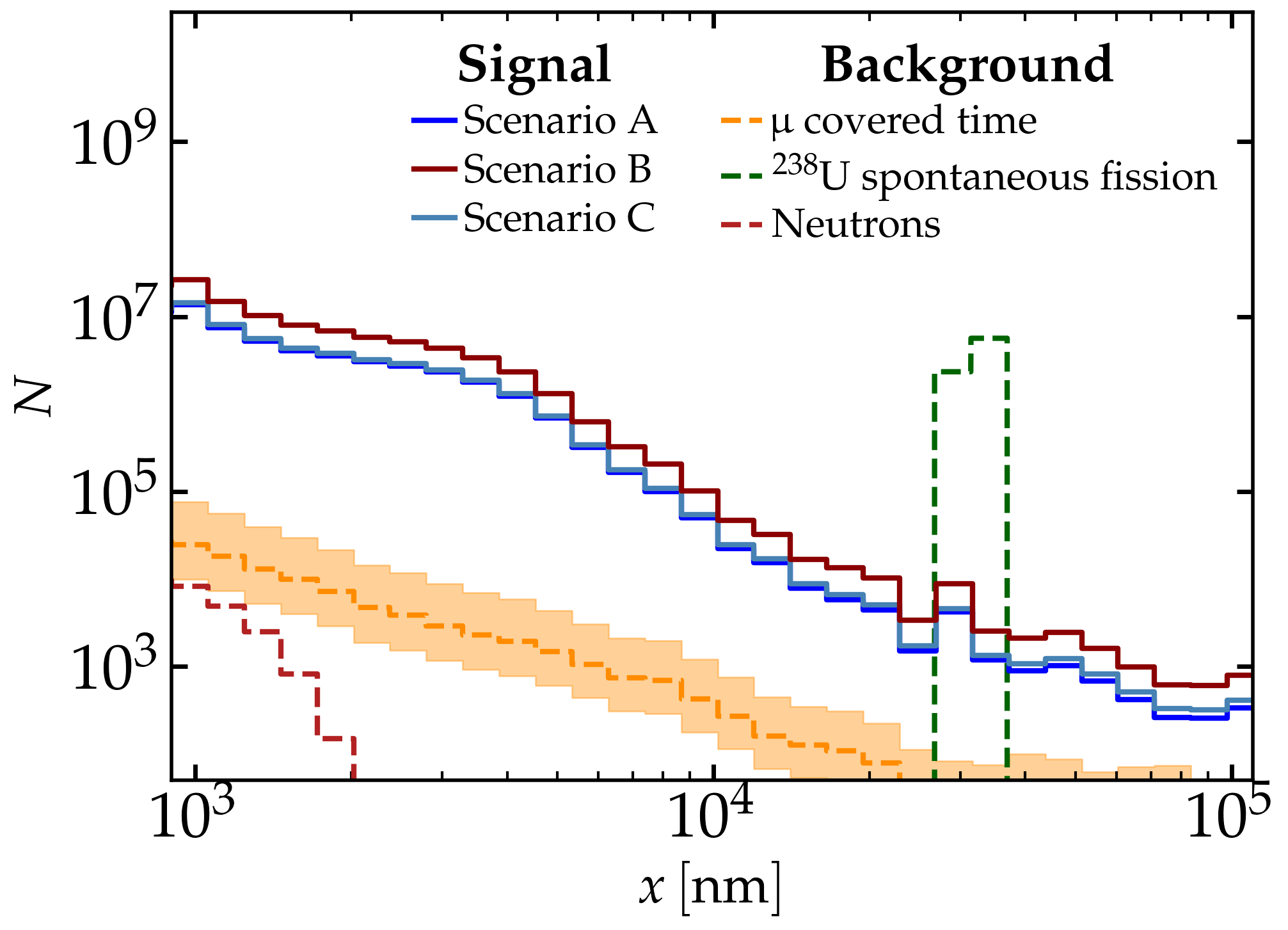} 
\caption{\label{fig:integralHalite} Same as Figure \ref{fig:integral}, but considering the continuous deposition of Halite after the sample is created and exposed. Top: considering a deposition rate of 30 m/kyr, bottom: considering instead a slower deposition rate of 2.5 m/kyr.}
\end{figure}

We considered also a more realistic scenario where the deposition of halite continues at a certain rate so that each sample is created, exposed under a shallow water overburden (a few tens of m, which is negligible in our simulation), and then shielded progressively by the subsequent halite deposits. The exact rate of deposition changes from different points in the Mediterranean basin, here we consider two extreme values of 30 m/kyr and 2.5 m/kyr, which were suggested in \cite{MEILIJSON2019374}. Results are shown in figure \ref{fig:integralHalite} where we suppose that the sample is exposed at the peak of the enhanced flux from the supernova. In a core sample, comparing the results at different depths would reveal an excess clearly ascribable to a transient event. We plan on following up in future works with dedicated simulations related to specific points of the Mediterranean where the deposition history is well known from existing sampling.

\section{Conclusions}

We have shown that paleo-detectors could be a powerful resource to investigate the past flux of cosmic rays, and perhaps the only way to access this interesting information. The collection and analysis of ${\sim}10\,$g samples from ${\sim}5\,$km deep underground to search for ${\sim}10\,\text{nm}-100\,$nm damages has been the basis of all previously suggested paleo-detectors applications. The collection of a similarly-sized sample of evaporite from the bottom of the Mediterranean Sea at a depth of a few km and a subsequent analysis in the search for tracks longer than a few $\upmu$m should be easier in almost all practical aspects. 

Halite has already been suggested as a good paleo-detector in previous studies. In this mineral, nuclear recoil tracks have already been observed and experimental efforts to identify the best track detection techniques are already ongoing ~\cite{BAUM2023101245, mastersthesis-full}.

The measurement of the flux of cosmic rays ${\sim}5.6\,$Myr ago would be an unprecedented result and would give important insights on the neighborhood of the Solar System and its evolution.

We note here that it has been suggested in \cite{Bouyahiaoui_2019} that the cosmic ray flux between $200\,\mathrm{GeV}$ and the so-called knee ($\simeq 4\,\mathrm{PeV}$) might be well explained with just two dominant contributors: the Vela supernova remnant, which is $\sim 11\,\mathrm{kyr}$ old and a local supernova $2-3\,\mathrm{Myr}$ old as the one expected from the ${}^{60}\mathrm{Fe}$ measurements. As both these events happened after the Messinian age, in this scenario one might expect that the evaporites produced in the Mediterranean desiccation were exposed to a significantly lower flux of CR than the one we measure today.

The search for SN signals during the Messinian period would be complementary to the ${}^{60}\mathrm{Fe}$ measurements, with comparable range of SN distance and investigated period. However, since our method does not involve unstable isotopes, it could also be performed on similar desiccation events from much older times.

Similar studies could be performed with minerals produced or projected during volcanic eruptions, exposed to the air, and then submerged by deposits from a subsequent eruption. The search for a suitable volcanic event, in particular in terms of the size of the deposits that need to be sufficient to shield the majority of secondary cosmic rays, is undergoing and will be the subject of future publications. In perspective, our goal would be to identify a series of events to sample the evolution of the cosmic ray flux in the past tens or possibly a few hundreds of Myr.

\begin{acknowledgments}
The authors thank the anonymous referees who greatly helped improve the manuscript with their comments. They also thank Marco Giammarchi, Max Stadelmaier, Antonio Condorelli, Barbara, Marco and Alessandro Caccianiga, Piera Ghia, and Olivier Deligny for the discussions and feedback received about this work. A great thank goes also to Alessandra Guglielmetti, Letizia Bonizzoni, and Cecilia Donini for introducing the authors to the detection of fission tracks in obsidian.
\end{acknowledgments}

%\bibliography{apssamp}{} 
%

\end{document}